\documentclass[twocolumn,aps,prb,amsmath,amssymb,floatfix,showpacs]{revtex4}
\usepackage{graphicx}
\begin{document}
\bibliographystyle{unsrt}

%TCIDATA{OutputFilter=LATEX.DLL}
%TCIDATA{Created=Sat Jul 13 21:35:27 2002}
%TCIDATA{LastRevised=Fri Jul 19 20:17:37 2002}
%TCIDATA{<META NAME="GraphicsSave" CONTENT="32">}
%TCIDATA{<META NAME="DocumentShell" CONTENT="Journal Articles\REVTeX - OSA Article">}
%TCIDATA{Language=American English}
%TCIDATA{CSTFile=revtxtci.cst}

\newtheorem{theorem}{Theorem}
\newtheorem{acknowledgement}[theorem]{Acknowledgement}
\newtheorem{algorithm}[theorem]{Algorithm}
\newtheorem{axiom}[theorem]{Axiom}
\newtheorem{claim}[theorem]{Claim}
\newtheorem{conclusion}[theorem]{Conclusion}
\newtheorem{condition}[theorem]{Condition}
\newtheorem{conjecture}[theorem]{Conjecture}
\newtheorem{corollary}[theorem]{Corollary}
\newtheorem{criterion}[theorem]{Criterion}
\newtheorem{definition}[theorem]{Definition}
\newtheorem{example}[theorem]{Example}
\newtheorem{exercise}[theorem]{Exercise}
\newtheorem{lemma}[theorem]{Lemma}
\newtheorem{notation}[theorem]{Notation}
\newtheorem{problem}[theorem]{Problem}
\newtheorem{proposition}[theorem]{Proposition}
\newtheorem{remark}[theorem]{Remark}
\newtheorem{solution}[theorem]{Solution}
\newtheorem{summary}[theorem]{Summary}

\title{Characterization of the $S = 9$ excited state in Fe$_8$Br$_8$ by Electron Paramagnetic Resonance}
\author{D. Zipse, J. M. North and N. S. Dalal}
\affiliation{Department of Chemistry and Biochemistry and National
High Magnetic Field Laboratory, Florida State University,
Tallahassee, Florida 32306}
\author{S. Hill and R. S. Edwards}
\affiliation{Department of Physics, University of Florida,
Gainesville, Florida 32611-8440}
\date{\today}

\begin{abstract}
High Frequency electron paramagnetic resonance has been used to
observe the magnetic dipole, $\Delta$ M$_s$ = $\pm$ 1, transitions
in the $S = 9$ excited state of the single molecule magnet
Fe$_8$Br$_8$. A Boltzmann analysis of the measured intensities
locates it at  24 $\pm$ 2 K above the $S = 10$ ground state, while
the line positions yield its magnetic parameters D = -0.27 K, E =
$\pm$0.05 K, and B$_4^0$ = -1.3$\times$ 10$^{-6}$ K. D is thus
smaller by 8\% and E larger by 7\% than for $S = 10$.  The
anisotropy barrier for $S = 9$ is estimated as 22 K,which is 25\%
smaller than that  for $S = 10$ (29 K). These data also help
assign the spin exchange constants(J's) and thus provide a basis
for improved electronic structure calculations of Fe$_8$Br$_8$.
\end{abstract}


\maketitle

\section{Introduction}
\begin{figure}
\includegraphics[width=.33\textwidth]{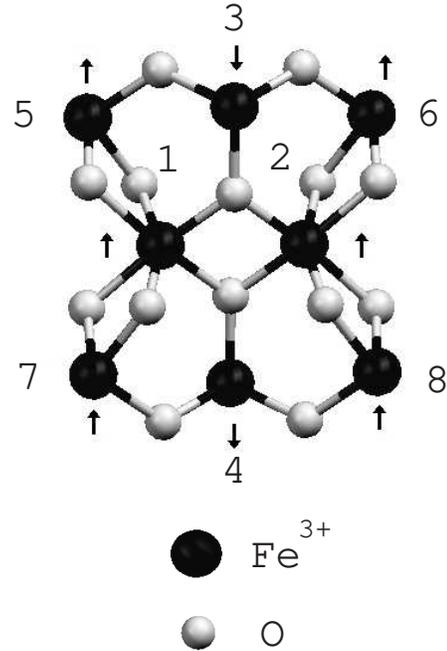}
\caption{Schematic representation of Fe$_8$Br$_8$ \cite{Delfs}.
The arrows represent spin orientations of the Fe$^{3+}$ ions in
the $S = 10$ ground state. The organic ligands and the eight
Br$^-$ anions have been omitted for clarity. Each Fe$^{3+}$ has $S
= 5/2$, thus the ground state can be seen as $S = 6 \times{5/2} -
2 \times{5/2} = 10$.} \label{struc}
\end{figure}

Single molecule magnets (SMM's), defined as compounds where a
magnetic domain can, in principle, be reduced to a single molecule
\cite{Sess1,Novak}, have recently been of high theoretical and
experimental interest due to their novel properties and potential
applications. These properties and potential applications include
quantum tunneling of their magnetization (QTM)
\cite{Fried,Ohm,Perenboom2,Mn12E}, whose detailed mechanism is
still not fully understood, molecular memory devices
\cite{Cheesman,Chud} and elements of quantum computers
\cite{Loss}. One of the best characterized SMM's is
{[(C$_6$H$_{15}$N$_3$)$_6$Fe$_8$($\mu_3$-O)$_2$($\mu_2$-OH)$_{12}$]Br$_7$(H$_2$O)}Br$\cdot$8H$_2$O,
abbreviated Fe$_{8}$Br$_{8}$ \cite{Weig,Barra}, whose main
spin-bearing skeleton is shown in Fig. \ref{struc}

Studies by magnetization \cite{Delfs}, neutron scattering
\cite{Caciuffo, Carretta}, electron paramagnetic resonance (EPR)
\cite{Barra,Perenboom,Macc1,Barra2,delBarco,Blinc,Park1,Park}, and
NMR \cite{Furu} techniques have established that the ground state
has a spin value $S = 10$. Figure \ref{struc} shows a schematic of
the spin configuration of the Fe$^{3+}$ ($S = 5/2$) ions. At
temperatures below 1 K, the magnetization relaxation takes places
via QTM \cite {Ohm}. While a great deal of progress has been made
in understanding the nature of QTM in Fe$_{8}$Br$_{8}$, many
questions still remain unclear. For example, the magnitudes of
calculated tunneling rates are much lower than the observed values
\cite{Wolf}. Secondly, there is a lack of data on the nature and
magnitude of the spin-exchange constants, J's, between the eight
Fe$^{3+}$ ions in the Fe$_8$Br$_8$ core. The best estimates come
from the temperature dependence of the \emph{dc} magnetic
susceptibility, $\chi$$_{dc}$ \cite{Delfs,Lasc}. The $\chi$$_{dc}$
\cite{Delfs} could be fitted by several sets of J's. One of the
criteria of such a procedure is the prediction  of the proper spin
$S$ value of the ground state, together with the location of the
excited states, in particular the $S = 9$ state. The $\chi_{dc}$
fit yielded at least two sets of J's, but the set providing the
better fit yielded the position of the $S = 9$ state to be less
than 0.5 cm$^{-1}$ above the ground state \cite{Delfs}. The other
set predicted the $S = 9$ state at greater than 25 cm$^{-1}$ (36
K) above the ground state. Clearly, from the point of view of
understanding the magnetic structure of Fe$_8$Br$_8$, it is
important to experimentally determine the location of the excited
states, in particular the $S = 9$ manifold. A peak around 30
cm$^{-1}$ has been reported in the far infrared spectrum
\cite{Mukhin1}, but it was not established whether it was a normal
molecular vibration involving metal ions, or a transition to the
$S = 9$ level from the $S = 10$ ground state. Moreover this peak
was not observed for the closely similar compound
Fe$_8$Br$_4$(ClO$_4$)$_4$, whose spin Hamiltonian (Eq. \ref{ham}),
parameters are known to be close to those of Fe$_{8}$Br$_{8}$
\cite{Barra3}.

The magnetic properties of the ground state of Fe$_8$Br$_8$ have
been well described by the following spin Hamiltonian, with $S =
10$
\cite{Barra,Perenboom,Macc1,Barra2,delBarco,Blinc,Park1,Park,Mukhin1,Barra3,EPR}:

\small
\begin{equation}
 \hat H = \mu _B \vec B\cdot
\vec{g}\cdot \vec S + DS_z^2 + E(S_x^2- S_y^2)+ B_4^0O_4^0 +
B_4^2O_4^2 + B_4^4O_4^4, \label{ham}
\end{equation}
\normalsize
\\[0.5cm]

where the first term is the Zeeman interaction, $D$ represents the
usual zero-field, uniaxial anisotropy parameter, and $E$ the
second order rhombic anisotropy. The fourth order terms are given
by: $O_4^0$ = 35 $S_z^4$ - [30 $S$($S$+1)- 25]$S_z^2$ - 6$S$($S$ +
1) + 3 $S^2$($S$ + 1)$^2$, $O_4^2$ = 1/4{[7$S_z^2$ - $S$($S$+1) -
5]($S_+^2$ + $S_-^2$)+ ($S_+^2$ + $S_-^2$)[7$S_z^2$ - $S$($S$+1) -
5]}, and $O_4^4$ = 1/2($S_+^4$ + $S_-^4$).

Here, we report on an EPR detection of the $S = 9$ state of
Fe$_{8}$Br$_{8}$, and preliminary evaluation of its spin
Hamiltonian parameters as defined in Eq. (\ref{ham}).  We observe
a series of peaks, called $\beta$ transitions, which we assign to
an $S = 9$ excited state of Fe$_8$Br$_8$. From the temperature
variation of the line intensities, we establish that the $S = 9$
manifold lies at 24 $\pm$ 2 K above the $S = 10$ ground state, in
contrast to the suggestion of $>$ 36 K from susceptibility
analysis \cite{Delfs}.

Section II below describes the single crystal preparation and the
EPR instrumentation. The results obtained, and their analysis, are
presented in section III, with the discussion presented in section
IV, and the main conclusions presented in section V.

\section{Experimental}
{[(C$_6$H$_{15}$N$_3$)$_6$Fe$_8$($\mu_3$-O)$_2$($\mu_2$-OH)$_{12}$]Br$_7$(H$_2$O)}Br$\cdot$8H$_2$O
(Fe$_{8}$Br$_{8}$) was synthesized following the procedure of
Weighardt \cite{Weig}. Relatively large (2 mm x 2 mm x 0.5 mm),
optical quality single crystals were prepared by slow evaporation.
The crystals were aligned with the Zeeman field applied along the
easy axis of magnetization by sight, to within a few degrees
\cite{Macc1}. The final orientation was confirmed by EPR
splittings, being the extremum for the canonical orientations. The
EPR measurements were made using a variable frequency (44 - 200
GHz), cavity based, high sensitivity spectrometer described
earlier \cite{Perenboom,Macc1,Mola}. The main component of the
spectrometer is a millimeterwave vector network analyzer (MVNA), a
phase-sensitive, fully sweepable, superheterodyne source-detection
system. A variable flow cryostat situated within the bore of an 17
T superconducting solenoid allows for temperatures down to 1.5 K,
with an accuracy of $\pm$ 0.01 K. The high sensitivity of the MVNA
technique (10$^9$ spins G$^{- 1}$ s$^{- 1}$) allows for
observation of the low level transitions of the Fe$_8$Br$_8$
ground and excited states in a single crystal, and for angular
variation studies, as described earlier
\cite{Park,Mn12E,Perenboom}.

\section{Results}

\begin{figure}
\includegraphics[width=.53\textwidth]{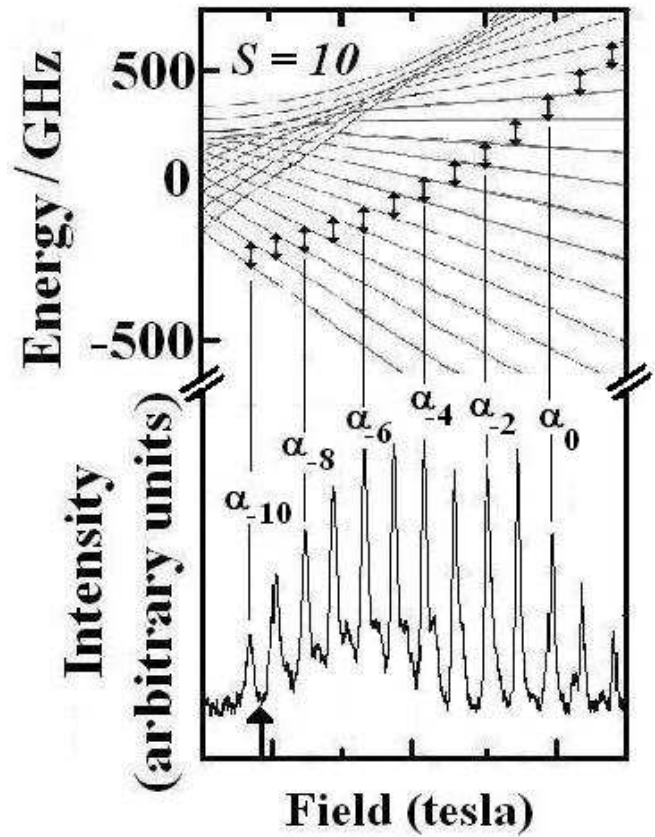}
\caption{EPR spectrum of Fe$_8$Br$_8$ at 131 GHz and 35 K with the
zeeman field applied along the easy axis (bottom panel), along
with the energy level diagram corresponding to the $S = 10$
 spin system (top panel). The $S = 10$ energy level
diagram has been constructed through the spin Hamiltonian
parameters of Caciuffo \emph{et al.} \cite{Caciuffo}.  The
single-headed arrow at 0.8 tesla designates the expected position
where a $\beta$$_{10}$ would appear if the $\beta$ transitions
originated from an $S = 10$ state.} \label{exp}
\end{figure}

Figure \ref{exp} (bottom panel) shows a typical EPR spectrum of
Fe$_8$Br$_8$ at 131 GHz with the Zeeman field applied along the
easy axis of an Fe$_8$Br$_8$ single crystal at 35 K. The spectrum
consists of a series of strong peaks $\alpha_{-10}$,
$\alpha_{-9}$, $\alpha_{-8}$, etc; the subscripts represent the
spin projection quantum number, M$_s$, corresponding to the level
from which the EPR absorption transition originates, in the $S =
10$ ground state following the convention introduced earlier
\cite{Barra,Macc1,Barra2}. The $\alpha$ transitions have been very
well analyzed earlier \cite{Barra,Macc1,Barra2}, and have been
shown to arise from the magnetic dipole ($\Delta$M$_s$ = $\pm$ 1)
transitions within the 21 M$_s$ levels of the $S = 10$ multiplet.
The specific peak assignment is indicated in the top panel for the
$\alpha$ transitions. In addition to the $\alpha$ resonances,
there are additional peaks present in Fig. \ref{exp} that are
labelled as the $\beta$ transitions and are the focus of the
current investigation.

Our analysis procedure consisted of three steps: (a) to ascertain
that the $\beta$ transitions are from an excited state, (b) to
determine the spin multiplicity of this excited state, and (c) to
deduce the spin Hamiltonian parameters for the $\beta$ spin system
and its energy position relative to the ground state ($S = 10$).

\begin{figure}
\includegraphics[width=.44\textwidth]{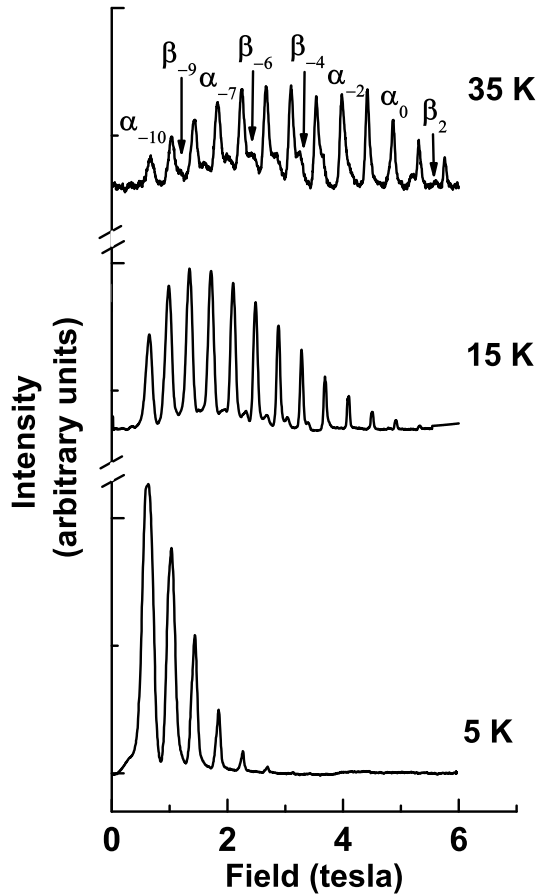}
\caption{Temperature dependence of EPR spectra of Fe$_8$Br$_8$,
for B $\parallel$ z at 131 GHz and 5, 15, and 35 K.}
\label{extemp}
\end{figure}

Direct evidence that the $\beta$ transitions originate from a
thermally populated excited state is provided by the temperature
dependence of their intensities. Figure \ref{extemp} shows
 spectra at 5, 15, and 35 K, respectively, giving
clear experimental evidence that the $\beta$ transitions arise
from a thermally populated excited state, because their
intensities rapidly decrease as the temperature is lowered.

\begin{figure}
\includegraphics[width=.44\textwidth]{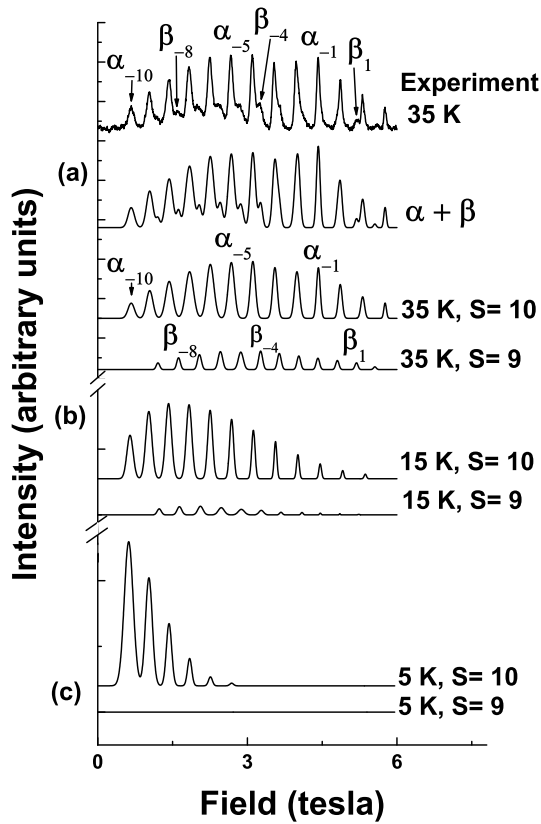}
\caption{(a) Experimental, resummed $\alpha$ and $\beta$,
separated $S = 9$, and separated $S = 10$ spectra at 35 K and 131
GHz for B $\parallel$ z. (b) Separated spectra at 15 K, (c) and at
5 K. } \label{tempdep}
\end{figure}

In order to quantitatively measure the intensities of the $\beta$
peaks as a function of temperature, the spectra of the ground and
excited states needed to be separated.  The separation was
accomplished by using a Gaussian fit for each individual peak. The
validity of  Gaussian fits, especially for the low field
transitions, has been previously established \cite{Park}.  The
criterion for the goodness of the fit was that the sum of the
$\alpha$ and $\beta$ transitions mirrors the experimental data
quite well by visual inspection and also by minimizing the
remaining intensity obtained by subtracting the Gaussian fits from
the experimental spectra. This remaining intensity was within the
experimental noise in our separation procedure. Each Gaussian fit
was then summed to yield a separated $S = 10$ spectrum. The
spectra were then separated by subtracting the Gaussian fits of
the $\alpha$ peaks from the experimental spectra. These Gaussian
fits were then summed, giving a separated $S = 10$ spectrum. The
remaining $\beta$ peaks were then fit with Gaussian functions, and
summed, yielding a separated $S = 9$ spectrum. We verified that
these separated spectra agreed well with the experimental data by
taking the sum of the $\alpha$ and $\beta$ spectra and comparing
it with experiment, as shown in Fig. \ref{tempdep}(a).

Figure \ref{tempdep}  also shows the relative decrease in the
intensity of the separated excited state ($\beta$) spectra in
relation to the $S = 10$ spectra as the temperature is decreased.
The spectral envelope present in the $S = 10$ spectra at 5, 15,
and 35 K is generally evident in the corresponding excited state
spectra, though the relative intensities are modified slightly due
to differing matrix elements in the transition probabilities,
P$_{nm}$ $\propto$ $ \mid  < \psi_n \mid S_+ \mid \psi_m >
\mid^2$, of the two spin systems, given by Eq. (\ref{element}):

\begin{equation}
P \propto [S \times (S + 1) - M_{s} \times (M_{s} + 1)].
\label{element}
\end{equation}

In order to analyze the intensity of the $\beta$ peaks we
normalized their intensities. The normalization process  involved
dividing the intensity of a $\beta$ peak, with a given M$_s$
value, by the corresponding $\alpha$ peak with the same M$_s$
value. This enabled us to ignore instrumental effects and spectral
envelope changes with temperature.  At 5 K, the $\beta$ peaks are
barely discernable from the noise level of the spectrum. At 10 K,
however, they increase in intensity enough to emerge from the
$\alpha$ transitions. The decreasing intensity of the $\beta$
peaks, upon lowering temperature, unambiguously designate the
$\beta$ peaks as being due to a thermally populated excited state.

Once it was concluded that the $\beta$ peaks originated from an
excited state, the spin multiplicity of the excited state needed
to be determined. The presence of an excited state (with perhaps
$S = 9$) close to the ground state of Fe$_8$Br$_8$ has been
previously inferred from magnetic susceptibility \cite{Delfs}, and
mentioned in subsequent $\mu$SR \cite{Lasc}, EPR \cite{Barra3},
and neutron diffraction studies \cite{Pontillon},  without any
evidence for its location or multiplicity. Herein, the spin
multiplicity of the excited state has been determined by two
independent methods. First, by the location of the leading peak
($\beta_{-9}$) in the set of peaks assigned to the $S = 9$ state.
Figure \ref{exp} shows an experimental spectrum taken at 131 GHz
and 35 K. As is evident in all spectra taken, there is no $\beta$
peak between the $\alpha_{-10}$ and $\alpha_{-9}$ peaks.
Furthermore, the $\alpha_{-10}$ transition is symmetric and shows
a clear Gaussian shape, as would be expected at this temperature
for a well separated, individual peak \cite{Park}. This lack of a
$\beta_{-10}$ transition, barring spin Hamiltonian parameters
being very different from those for the ground state, is strong
evidence that the excited state is $S = 9$. The location of the
first $\beta$ peak is consistent with slightly modified spin
Hamiltonian parameters and a spin multiplicity of $S = 9$ for the
excited state, as anticipated theoretically \cite{Delfs}.

\begin{table}
\caption{Experimental and simulated peak positions at 131 GHz for
the $S = 9$ state, for B $\parallel$ z, using D = -0.27 K, E =
$\pm$0.05 ($\pm$0.015) K, B$_4$$^0$ = -1.3 $\times$ 10$^{-6}$ K
,B$_4$$^2$ = 0 K, B$_4$$^4$ = 0 K   }
\begin{tabular}{|l|l|l|}
\hline\hline

\textbf{Transition} &\textbf{Experiment}& \textbf{Simulation} \\

M$_s$$\to$ M$_{s+1}$ && \\ & (tesla)& (tesla)  \\

\hline\hline $-9$ $\to$ $-8$, $\beta$$_{-9}$ & 1.2082 & 1.2282\\
\hline $-8$ $\to$ $-7$, $\beta$$_{-8}$ & 1.6208 & 1.6400\\
\hline $-7$ $\to$ $-6$, $\beta$$_{-7}$ & 2.0409 & 2.0608\\
\hline $-6$ $\to$ $-5$, $\beta$$_{-6}$ & 2.4610 & 2.4691   \\
\hline $-5$ $\to$ $-4$, $\beta$$_{-5}$ & 2.8679 & 2.8727\\
\hline $-4$ $\to$ $-3$, $\beta$$_{-4}$ &3.2659& 3.2760\\
\hline $-3$ $\to$ $-2$, $\beta$$_{-3}$ &3.6394&3.6721\\
\hline $-2$ $\to$ $-1$, $\beta$$_{-2}$ &4.0312&4.0585\\
\hline $-1$ $\to$ $0$, $\beta$$_{-1}$ &4.4126 &4.4571\\
\hline

\end{tabular}

\end{table}
\begin{figure}
\includegraphics[width=.44\textwidth]{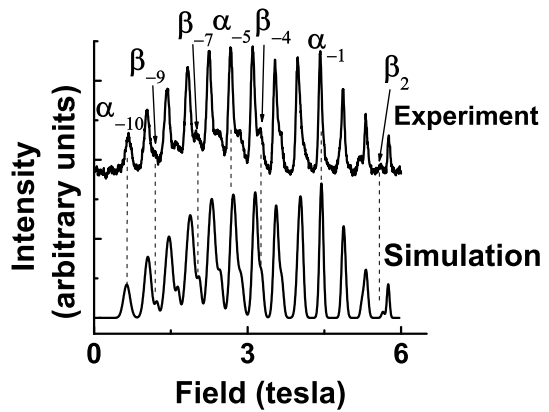}
\caption{Experimental and simulated EPR spectra of Fe$_8$Br$_8$,
for B $\parallel$ z at 35 K and 131 GHz. } \label{sim}
\end{figure}
Additional, more quantitative support, that the spin  $S$ of this
excited state is indeed $S = 9$ was provided by computer
simulations which were run using SIM \cite{SIM}. The procedure was
first checked for the $S = 10$ state for which the parameters are
known \cite{Caciuffo,Barra2,Mukhin1}. The simulations for the $S =
10$ state were performed with the spin Hamiltonian parameters
previously determined by Caciuffo \emph{et al.} \cite{Caciuffo}
using neutron scattering. A typical comparison is shown in Fig.
\ref{sim}. The simulated spectra for the $S =10 $ ground state
were in close agreement with our experimental results, thereby
validating the simulation procedure.

\begin{figure}
\includegraphics[width=.44\textwidth]{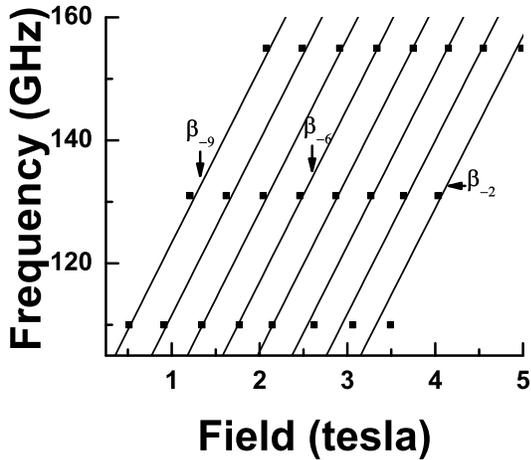}
\caption{Frequency dependence of the first eight $\beta$
transitions in Fe$_8$Br$_8$ with B $\parallel$ z. The solid lines
portray the frequency dependence given by the spin Hamiltonian
parameters determined in this work, whereas the solid squares are
the observed field positions.} \label{fvf}
\end{figure}

\begin{figure}
\includegraphics[width=.44\textwidth]{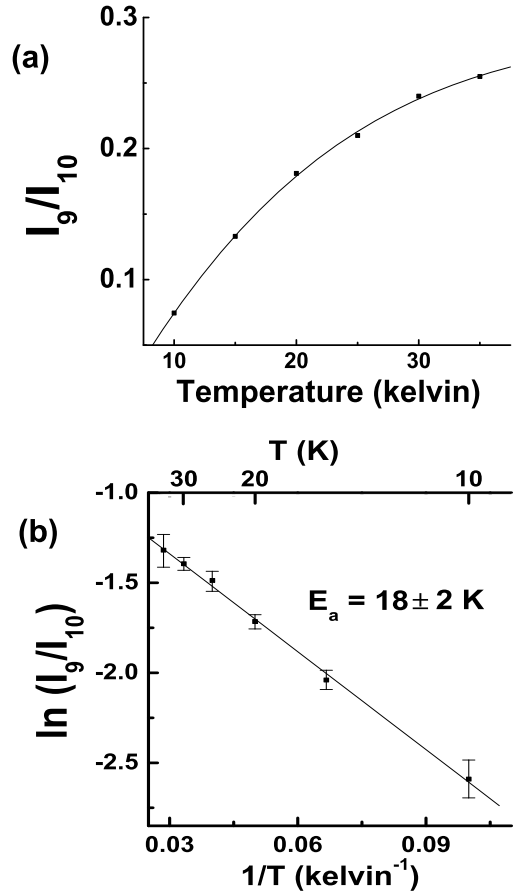}
\caption{(a) Temperature Dependence of the intensity of the -8 to
-7 transition in the $S = 9$ state, I$_9$, normalized to the -8 to
-7 transition in the $S = 10$ state, I$_{10}$. The curve joining
the experimental points is a guide to the eye. (b) Boltzmann
analysis of the normalized intensities of the $S = 9$ spin state.
The slope yields the excitation energy as 18 $\pm$ 2 K between
level with the same M$_s$ values in the $S = 10$ and $S = 9$
manifolds. Addition of the zero field splitting between the M$_s =
-10$ and M$_s = -9$ levels (5.5 K) leads to the location of the $S
= 9$ state at 24 $\pm$ 2 K above the ground state.}
\label{arrhenius}
\end{figure}
The $\beta$ transitions were then accurately simulated for the
three frequencies utilized (110, 131, and 155 GHz), using the spin
Hamiltonian parameters obtained in the present study. A
field-frequency plot is shown in Fig. \ref{fvf}. Quite good
agreement can be seen between the simulated curves (continuous
lines) and the observed peak positions. A more explicit and
quantitative comparison is shown in Table I, using the measured
data for 131 GHz. The D parameter decreases in magnitude by 8\%,
to -0.27 K, while the E term increases by 7\%, to $\pm$ -0.05 K
($\pm$ 0.015 K). Surprisingly, the B$_4^0$ term is similar in
magnitude, -1.3 $\times$ 10$^{-6}$ K, but opposite in sign to the
$S = 10$ parameter (1.01 $\times$ 10$^{-6}$ K). Due to the fact
that the experimental spectra were taken along the easy axis of
magnetization, the B$_4^2$ and B$_4^4$ terms were not included in
the simulations. A simulated spectrum at 35 K, with the spin
Hamiltonian parameters determined for both the $S = 10$ and $S =
9$ states is shown in Fig. \ref{sim}. Linewidths were not an
optimized parameter in the simulations, though the relative
intensities matched the experimental data well. Therefore, the
simulated spectrum presented in this work is the sum of the $S =
10$ and $S = 9$ separated spectra using peak positions generated
by SIM \cite{SIM}. The agreement between the experimental and
simulated spectra, as shown in Fig. \ref{sim}, can be seen to be
quite satisfactory, thereby supporting the parameter assignment.
We are thus able to assign the full experimental spectrum to
transitions in the $S = 10$ multiplet ($\alpha$'s) and the $S = 9$
multiplet ($\beta$'s), as shown in Figure \ref{exp}.

Once the spin Hamiltonian parameters had been determined, the
relative intensities of the $\alpha$ and $\beta$ transitions were
used, at temperatures from 5 to 35 K, to determine the location of
the $S = 9$ state above the ground state. The intensity of a
specific peak between given M$_s$ and M$_{s+1}$ states, is
proportional to the population difference between the M$_s$ and
M$_{s+1}$ states and the transition probability, $P$, as given in
Eq. (\ref{popdiff}):

\begin{equation}
I \propto P (N_{M_s} - N_{M_{s+1}})/Z. \label{popdiff}
\end{equation}

Therefore, assuming very similar partition functions and D values
for the two states, the intensity ratio between two transitions of
the same M$_s$ states in the $S = 9$ and $S = 10$ manifolds,
respectively, is given by Eq. (\ref{ratio}):

\begin{equation}
I_9/I_{10} = (P_9/P_{10}) exp (-\Delta E_{10-9}/kT),
\label{ratio}
\end{equation}

where k is the Boltzmann constant and $\Delta$E$_{10-9}$ is the
energy difference between given M$_s$ states in the $S = 10$ and
$S = 9$ manifolds. The areas of the Gaussian fits, for a given
M$_s$ to M$_{s+1}$ transition, were factored by their transition
probabilities, in both the $S = 9$ and $S = 10$ manifolds, in
order to determine their ratios. Due to the fact that each $\beta$
transition is normalized to its corresponding $\alpha$ transition,
the ratio of intensities should be constant, regardless of the
specific M$_s$ pair, for any given temperature.

\begin{figure}
\includegraphics[width=.44\textwidth]{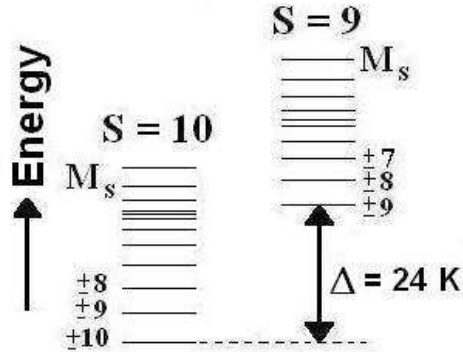}
\caption{Schematic for the energy levels of both the $S = 10$ and
the $S = 9$ states in zero magnetic field. $S = 9$ is located at
an energy $\Delta$ = 24 $\pm$ 2 K as marked.} \label{10-9ELD}
\end{figure}

A Boltzmann analysis of the intensity ratios is shown in Fig.
\ref{arrhenius}(b). The ratios of intensities of the $\beta$ to
$\alpha$ transitions were compared at 5, 15, 20, 25, 30 and 35 K,
and plotted vs. inverse temperature (T$^{-1}$). The slope yields
the energy difference (18 K) between a given M$_s$(9) state and
the corresponding M$_s$(10) state. Therefore, the energy
difference between the M$_s$ = 10 and M$_s$ = 9 in zero field
(5.46 K) must be added to the energy obtained from the Boltzmann
analysis, yielding an energy difference $\Delta$ of 24 $\pm$ 2 K
(16.7 $\pm$ 1.5 cm$^{-1}$). Figure \ref{10-9ELD} shows a schematic
of the energy levels in zero field for the $S = 10$ and $S = 9$
manifolds based on the present study. Clearly, the higher M$_s$
levels of the $S = 9$ state overlap with the lower M$_s$ levels of
the $S = 10$ state, indicating at least a partial breakdown of the
single spin model.

\section{Discussion}
Our determination that the $S = 9$ manifold in Fe$_8$Br$_8$ is
located 24 $\pm$ 2 K above the $S = 10$ ground state is in
contrast to with earlier suggestions based on magnetic
susceptibility ($>$ 36 K) \cite{Delfs}. The four central Fe$^{3+}$
ions in the Fe$_8$Br$_8$ core (Fe$_1$$^{3+}$, Fe$_2$$^{3+}$,
Fe$_3$$^{3+}$, and Fe$_4$$^{3+}$ (as shown in Fig. 1)) can be
described as that of a butterfly configuration. All magnetic
coupling interactions, J, between Fe$^{3+}$ ions in Fe$_8$Br$_8$
have been determined to be antiferromagnetic \cite{Delfs}.
Therefore, the magnitude of these coupling constants dictates both
the location and spin topology of the $S = 9$ excited state.

 As mentioned in the Introduction, a detailed analysis of the \emph{dc} magnetic
susceptibility data led Delfs \emph{et al.} \cite {Delfs} to two
reasonable sets of exchange parameters; (a) J$_{1-2}$ = 20
cm$^{-1}$, J$_{1-3}$ = 120 cm$^{-1}$, J$_{1-5}$ = 15 cm$^{-1}$,
and J$_{3-5}$ = 35 cm$^{-1}$, and (b) J$_{1-2}$ = 102 cm$^{-1}$,
J$_{1-3}$ = 120 cm$^{-1}$, J$_{1-5}$ = 15 cm$^{-1}$, and J$_{3-5}$
= 35 cm$^{-1}$. While set (b) provided a much better fit to the
experimental data, it predicted the position of the first excited
state, $S = 9$, at less than 0.5 cm$^{-1}$ above the $S = 10$
ground state. We do note, however, that Delfs $\emph{et al.}$ did
not include any zero field splitting terms in their susceptibility
analysis. Nevertheless, this same basic configuration of exchange
constants has been recently supported by detailed symmetry-based
calculations by Raghu $\emph{et al.}$ \cite{Raghu}.  Though the
magnitudes of the coupling constants calculated by these authors
\cite{Raghu} are different from those of Delfs \cite{Delfs}, the
dominance of the J$_{1-3}$ interaction over other magnetic
couplings remains consistent. The present study supports the
essential correctness of set (a).

The coupling set (a) of Delfs \emph{et al.} \cite{Delfs} and the
best set of Raghu \emph{et al.} \cite{Raghu}, both show that
J$_{1-3}$ dominates the exchange interactions. The perturbation
leading to the $S = 9$ excited state must result from the smallest
difference in J's acting on the same ion or symmetrically
equivalent set  of Fe$^{3+}$ ions. Thus it seems reasonable to
deduce that this perturbation does not involve J$_{1-3}$, hence
the butterfly core (Fe$_1$, Fe$_2$, Fe$_3$, Fe$_4$), but rather
the Fe's on the corners of the cluster (Fe$_5$, Fe$_6$, Fe$_7$,
Fe$_8$) via some linear combination of their wavefunctions.

The spin Hamiltonian parameters determined for the $S = 9$ state
provide some insight to the origin of the anisotropy of the
cluster. The $S = 9$  parameters  are slightly different from
those of the $S = 10$ ground state.  The 7$\%$ larger E value for
$S = 9$ is in accord with increased transverse distortion in the
Fe$_8$ structure.  The decrease in D with decreasing magnetic
moment indicates that the anisotropy present in the Fe$_8$Br$_8$
core has some dipolar contribution rather than arising purely from
a spin-orbit interaction. Similarly, the change in sign of the
B$_4^0$ term indicates that B$_4^0$ originates from many-body
interactions between Fe$^{3+}$ ions, and not from a collective sum
of individual B$_4^0$ terms. This argument is in line with the
fact that a B$_4$ term needs an effective interaction involving at
least four spins. Alternatively, this significant change in
B$_4^0$ may be a result of the breakdown of the single spin model,
as has been proposed by Katsnelson \emph{et al.} in connection
with Mn$_{12}$-Acetate \cite{Harmon}. Additional, detailed angular
variation studies are underway for precise measurement and
understanding of these questions.

The anisotropy barrier, estimated from D and E values for the $S =
9$ manifold is 22 K, as compared  to that for the $S = 10$ ground
state (29 K).

\section{Conclusions}
Using variable frequency, high-field, EPR measurements on single
crystals of Fe$_{8}$Br$_{8}$, we have detected a set of
transitions, labelled as $\beta_i$, which have been conclusively
assigned to the $S = 9$ spin multiplet, located at an energy
$\Delta$ = 24 $\pm$ 2 K (17 $\pm 1.5$ cm$^{-1}$ ) above the ground
($S = 10$) state. The spin Hamiltonian parameters have been
determined to a good accuracy, and  differ  from those of the $S =
10$ state: D is smaller by 8\%, while E is larger by 7\%. These
parameters yield the anisotropy barrier ($\sim$ DS$_z$$^2$ $\sim$
22 K), about 25\% smaller than for $S = 10$. B$_4$$^0$ for $S = 9$
also shows a dramatic change, the sign is opposite to that for the
$S = 10$ state. Although electronic structure calculations have
been reported for Fe$_{8}$Br$_{8}$ \cite{Raghu,Kortus,Pederson},
there has been little definitive data on excited states. The
results of the present study should serve as sensitive basis for
more refined theoretical modelling of the bonding and magnetic
properties of these materials.

\section{Acknowledgement}
We would like to thank the National Science Foundation (NIRT Grant
No. DMR 0103290) for financial support.

\end{document}